\documentclass[letterpaper, 10 pt, conference]{ieeeconf}
\usepackage{cite}
\usepackage{amsmath,amssymb,amsfonts}
\usepackage{algorithmic}

\usepackage{textcomp}
\usepackage{xcolor}

\usepackage[pdftex]{graphicx}
\usepackage{arydshln}
\usepackage{subcaption}
\usepackage{graphicx}
\usepackage{float}
\usepackage{multirow}
\usepackage{eucal}
\usepackage{comment}
\usepackage{mathtools}

\IEEEoverridecommandlockouts                              
\usepackage{ntheorem}
\usepackage{amsmath}
\usepackage{amsfonts}

\title{\LARGE \bf
Unraveling the Neural Network: Identifying Temporal Labeling of Visual Events through EEG-Based Functional Connectivity Analysis of Brain Regions
}

\author{Sina khoonbani and Hasan Ramezanian
\thanks{This work was not supported by any organization}
\thanks{s.khoonbani is with the Department of mechanical Engineering , Amirkabir University of tecnology( Tehran polytechnic), Tehran, Iran
        {\tt\small khoonbani@aut.ac.ir}}%
\thanks{H. Ramezanian is with the Department of Mechanical Engineering, Amirkabir University of Technology,
        Tehran, Iran
        {\tt\small hramezanian@aut.ac.ir}}%
}

\begin{document}

\maketitle
\thispagestyle{empty}
\pagestyle{empty}

\begin{abstract}
Understanding the complex interplay between the brain and a dynamic environment necessitates the continuous generation and updating of expectations for forthcoming events and their corresponding sensory and motor responses. This study investigates the interconnectivity patterns associated with time perception in predictable and unpredictable conditions. EEG signals were obtained from an existing database, encompassing an experiment conducted on healthy participants subjected to two conditions: predictable and unpredictable, across various time delays. Functional connectivity between brain regions was estimated using the phase lag index method, allowing for the identification of differences in time perception between conditions. Comparative analysis revealed significant variations, particularly in the gamma, beta, and theta frequency bands, with more pronounced differences observed in the predictable condition. Subsequent exploration of the dissimilarities within each delay demonstrated significant differences across all delays. Notably, the unpredictable condition exhibited increased connectivity within the alpha band during the 400-ms delay, specifically between occipital and temporal regions, with higher mean connectivity compared to the predictable condition. In the delta band, distinct connectivity patterns emerged, involving connections between central and frontal regions across different delays. Notably, heightened connectivity between central and prefrontal regions was observed during the 83-ms delay. The right hemisphere of the prefrontal cortex played a pivotal role in time perception. Furthermore, a decline in connectivity across the delta, theta, and beta bands was observed during the longest delay (800 ms) in both conditions, relative to other delays. These findings enhance our understanding of the neural mechanisms underlying time perception and underscore the impact of predictability on connectivity dynamics.

\end{abstract}

{\small \vspace{0.3 cm}\textit{Keywords:} \textbf{temporal perception, predictable events, unpredictable events, EEG signals, functional connectivity analysis, Phase Lag Index (PLI), gamma band, beta band, theta band, alpha band, delta band, frontal lobe}}


\section{Introduction}
\label{sec:intro}
The human brain can be regarded as a complex structure composed of numerous interconnected networks. As the brain functions as an integrated system, the performance of specific tasks within the brain is not solely derived from the isolated activities of its regions, but rather relies on the interaction and communication among its constituent parts. Therefore, examining the connections and interactions among the brain regions is crucial for a comprehensive understanding of brain activity. Investigations of brain interactions are conducted at various levels. Considering that the brain operates at a high speed, although the physical connections between regions may remain stable for several seconds, the functional communication (functional connectivity) between activities of different regions can vary on a small time scale, even within milliseconds.

In this study, electroencephalography (EEG) signals are utilized to record the electrical activity of the brain\cite{yelamanchili2018neural}. EEG signals provide an approximate measurement of postsynaptic activity of neuronal cells with a temporal resolution on the order of milliseconds, enabling the description of brain activity dynamics \cite{barttfeld2012state}. Hence, the use of EEG can be valuable in investigating functional connections between regions due to its high temporal resolution. Various estimators, such as correlation estimators, partial coherence, mutual information, phase locking value, phase lag index, phase slope index, weighted phase lag index, etc., have been employed in studies to explore functional connections \cite{cover1999elements,van2014functional,vinck2011improved}.
The concept of time is defined as a quantitative measure of various sequential events to compare their duration or the interval between them \cite{dictionary2016new, minnoye2022personalized, tajdari2022flow, tajdari2023online, tajdari2022onlinearXiv, tajdari20212d, tajdari20234d, tajdari2022adaptive, paydarfar2020investigation, tajdari2022next, tajdari2023optimal, tajdari2023non, tajdari2020feedback, paydarfar2023serum, tajdari2017switching, tajdari2021adaptive, tajdari2017design, tajdari2020semi, tajdari2019integrated}. The traditional view, influenced by St. Augustine of Hippo philosophy, has been of interest to cognitive neuroscientists, suggesting that time is a composite of simultaneous and disjointed mental concepts. For instance, the past is constructed through memories, the present is shaped by attention, and the future is anticipated through prediction \cite{khakzand2016framework}. To interact with the changing environment we live in, the brain constantly requires generating and updating predictions about future events \cite{gallistel2000time}. In fact, the ability to predict the timing of events in the environment enables humans to allocate the necessary processes for comprehension and appropriate action \cite{pasquereau2015dopamine}. Additionally, in order to communicate with the surrounding environment, the brain needs to perceive the timing of events and utilize estimates of time to regulate sensory and motor responses. However, it is still unclear how the brain processes time. Furthermore, it has been well-established that the temporal concept is correlated with perceptual processing \cite{cravo2013temporal,lakatos2008entrainment,rohenkohl2014combining, ghaffari2018new, khodayari2015new, rad2016design, rad2015hysteresis}.
Recognizing regular timing is not the only temporal pattern that the brain is capable of perceiving. Recent studies on humans and primates have focused on investigating the brain's capacity to identify more complex temporal distributions apart from regular intervals \cite{janssen2005representation,jazayeri2010temporal}. To understand how the brain operates during different time periods and utilizes them for external interactions, it is necessary to identify the regions and mechanisms involved in the temporal processing. It is evident that the brain should not rely solely on one mechanism or a specific region for all different temporal ranges. While an internal clock, which perceives time in the form of 24-hour intervals, has been identified as a known system \cite{darlington1998closing}, neural mechanisms operating at sub-millisecond to second timescales, representing temporal intervals, are still not fully comprehended \cite{muller2014perceiving}.
Cellular activity changes related to temporal processing have been observed in the cerebellum \cite{perrett1998temporal}, thalamus \cite{tanaka2007cognitive}, posterior parietal cortex \cite{leon2003representation}, prefrontal cortex \cite{brody2003timing,genovesio2006neuronal,genovesio2009feature,oshio2008temporal} and motor cortex \cite{lebedev2008decoding} in monkey behavior. Furthermore, neurological disorders ultimately alter the functioning of the cerebral cortex, which may contribute to other timing impairments in Parkinson's Disease and early stages of diseases such as Huntington. Specifically, damage to the right frontal cortex and subcortical structures disrupts temporal perception \cite{harrington1998cortical}. Findings suggest that the prefrontal cortical regions may control attention and memory, which interact with timing processes \cite{matell2005not}.
In study \cite{samaha2015top}, the aim was to investigate whether the prediction of stimulus timing modulates sensory labeling accuracy and whether the timing between the cue and stimulus has an impact on this modulation. EEG signals were recorded in predictable block conditions with stimuli presented after a fixed delay relative to the cue, and in unpredictable block conditions with stimuli presented after variable delay sequences (83, 150,400, and 800 milliseconds) relative to the visual cue. The predictability effect on behavior was observed at posterior parietal sites in the alpha power, specifically during the longest delay period (800 milliseconds). In fact, a significant difference and improvement in performance were observed between the predictable and unpredictable conditions at the 800-millisecond delay, indicating the influence of predictability on sensory processing.
To better understand temporal processing in the brain, functional connectivity analyses were conducted using EEG signals and the Phase Lag Index (PLI) as a measure of connectivity in different brain regions.
\section{Materials and Methods}
\subsection{Data Introduction}
In this study, data recorded from the article \cite{samaha2015top} were utilized. The experiment involved EEG data recorded at a sampling frequency of 1000 Hz with 64 AGCL/AG electrodes referenced to the mastoid bone. The impedance of the channels was maintained below 6 kilohms.

\begin{figure}[tb]
	\centering
	\includegraphics[width= 1\linewidth]{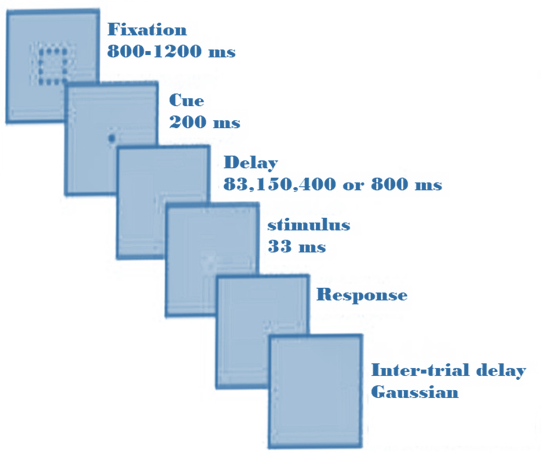}
	\caption{General experimental protocol\cite{samaha2015top}}
	\label{fig.1}
\end{figure}

The experiment was conducted on 29 healthy male individuals with normal vision, right-handed, and with an average age of 24 years. The experimental protocol was approved by the ethical committee of sirjan University of Medical and Sciences.
A 24-inch display screen with a resolution of 1920×1024 pixels was used in the experiment. The general experimental protocol is illustrated in Fig. \ref{fig.1}. Initially, to ensure the participant's fixation and concentration, a fixation cross appeared. Then, stimuli were presented for 200 milliseconds, followed by a delay period. Subsequently, a visual stimulus in the form of clockwise or counterclockwise rotation appeared, and the participant was required to press the right or left key, respectively, depending on the rotation direction. Following that, to introduce a non-rhythmic aspect to the experiment (since rhythmic experiments involve the presentation of stimuli at brain frequencies), a delay was introduced. The delay was randomly sampled from a Gaussian \cite{tajdari2021image, tajdari2017robust, tajdari2020intelligent, tajdari2019fuzzy, tajdari2022feature, tajdari2022implementation, tajdari2021discrete, tajdari2020intelligentcontrol, tajdari2021simultaneous} distribution with a mean of 900 milliseconds and a variance of 600 milliseconds.

As mentioned, prior to stimulus presentation, a delay was introduced to transform the experiment into two conditions: predictable and unpredictable. Each block was predictable. In the predictable or unpredictable, it included 48 trials, where only one of the delays (83, 150,400, or 800 milliseconds) was considered as the target delay between the cue and the stimulus for all trials in each experiment. In the unpredictable condition, for each trial, the delay was randomly selected from the four delays (83, 150,400, or 800 milliseconds) to ensure that no delay coefficient was a multiple of another, thus avoiding the creation of harmonics in a specific rhythm. Additionally, the timing of each frame of the monitor (60 Hz frequency) was taken into account when selecting these delays to ensure proper alignment. The recorded signal aimed to investigate time perception in humans and the role of brain oscillations in timing information processing.
In their study, they concluded that the temporal prediction effect on alpha power exists during the longest delay period (800 milliseconds). They observed that after the cue presentation, alpha power decreased in the predictable condition (from 172 to 373 milliseconds after the cue), compared to the unpredictable condition. However, before the stimulus presentation, alpha power increased in the predictable condition (305 milliseconds before the stimulus), compared to the unpredictable condition. The significant difference and improvement in performance were only observed during the longest delay period (800 milliseconds) between the two conditions\cite{samaha2015top}.

\subsection{Preprocessing}
\begin{figure}[tb]
	\centering
	\includegraphics[width= 1\linewidth]{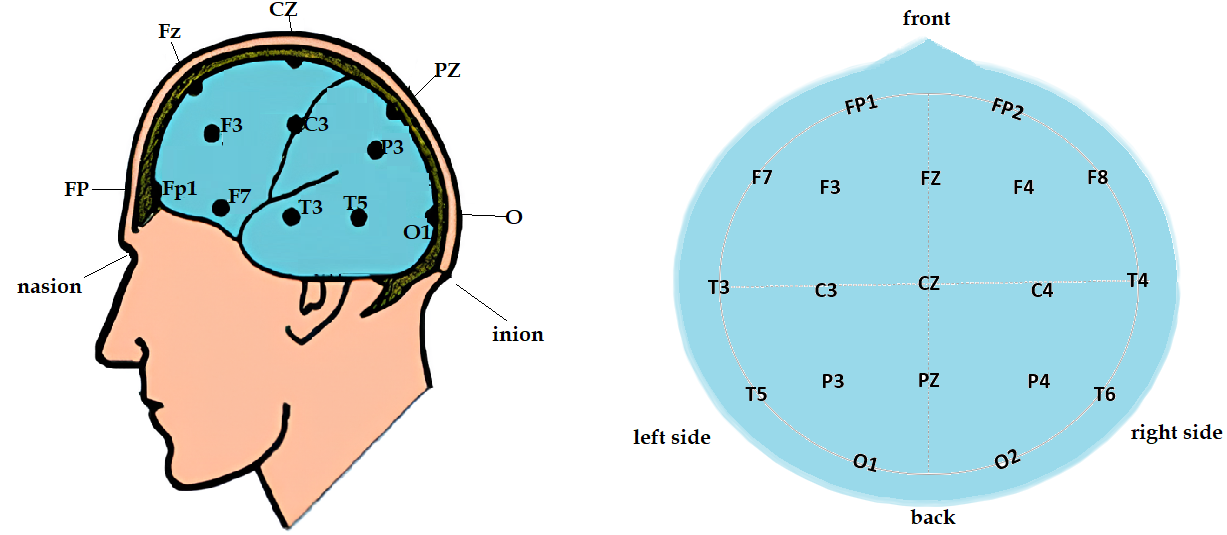}
	\caption{19 selected electrodes}
	\label{fig.2}
\end{figure}

The preprocessing steps were performed using MATLAB software and the EEGLAB toolbox. Initially, the recorded signals were imported into the software, and their DC offset was removed using a line base removal technique. To eliminate drifts and artifacts caused by very low-frequency components with significantly higher magnitude and power compared to EEG signals, a high-pass filter with a cutoff frequency of 0.5 Hz was applied.
After filtering, visual inspection was performed on the signal, and the channels that were corrupted were replaced with neighboring channels using healthy channel signals. Subsequently, relative and visual data interpolation were applied to clean the data. The purpose of this process was to eliminate common noise present in all channels. For further cleaning, Independent Component Analysis (ICA) algorithm was applied to the signals, and then artifacts related to blinking, eye and neck movement, and so on, were removed.
From the remaining 62 electrodes, a subset of 19 standard 10-20 electrodes fig.\ref{fig.2}  was selected for further analysis. These electrodes were chosen based on clinical relevance and their ability to cover the entire brain.

After the electrode selection, segmentation was performed based on the specified timing as described in reference \cite{samaha2015top}. In this segmentation, a time interval of 500 milliseconds prior to the onset of the stimulus to 500 milliseconds after the stimulation was chosen.

\subsection{estimate of correlation}
In the present study, the PLI (Phase Lag Index) was utilized to investigate the phase synchronization relationships in brain oscillations during a state where the phases of two oscillators were coupled. Phase synchronization refers to the simultaneous occurrence of phases despite their non-coherence in amplitude \cite{rosenblum1996phase, tajdari2022optimal, tajdari2022dynamic}. The foundations and theory of this method have been further examined in the following sections.
\subsubsection{Phase lag index}
The Phase Lag Index (PLI) is a functional connectivity estimator defined by the equation\ref{eq:example1}, which, theoretically, is resistant to volume conduction artifacts \cite{stam2007phase}.

\begin{equation}
	\centering
	\includegraphics[width= 0.3\linewidth]{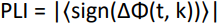}
	\label{eq:example1}
\end{equation}

In this equation, "sign" denotes the sign function,$\Phi_x$ and $\Phi_Y$ represent the phase of the momentary samples of two time series N and k = 1...N at discrete time steps tk, $\Delta\Phi = \Phi_x - \Phi_y$, and N refers to the number of samples. Additionally, PLI values range from 0 to 1, where PLI is a statistical measure of dependence between time series that reflects the strength of their coupling. This method has been employed due to its lower sensitivity to volume conduction and common sources \cite{van2015opportunities}.
In this research, functional connectivity related to each of the 29 participants was estimated in the following frequency bands: delta (1-4 Hz), theta (4-8 Hz), alpha (8-13 Hz), beta (13-30 Hz), and gamma (30-49 Hz). The obtained results form 40 symmetric 19x19 matrixs, where the rows and columns correspond to the channels used in this study. Each element of the matrix represents the connectivity between two recorded channels. The elements of the matrix are symmetric with respect to the main diagonal because directionality is not relevant in functional connectivity. Therefore, in each matrix, the number of different states is equal to the following value: 

\begin{equation}
	\centering
	\includegraphics[width= 0.4\linewidth]{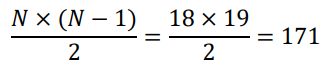}
	\label{eq:example}
\end{equation}
The functional relationships were estimated by calculating the average values of PLI. Similarly, the correlation matrix was obtained for both predictable and unpredictable states, encompassing all possible delays.
\subsubsection{statistical analysis of data}

After applying the necessary preprocessing steps to the data and estimating all the connections between brain channels, statistical analyses were performed. 
In this study, SPSS24 was used for performing the statistical analyses. Initially, the normality of the data was ensured using the Kolmogorov-Smirnov test with

$P\text{-Value} \geq 0.05$  . The first step involved comparing the clockwise and counterclockwise stimulations at each of the four delays to determine if there was a significant difference. For this purpose, paired t-tests were conducted with eight paired samples (individuals in two different conditions). Subsequently, to examine whether there were significant differences among different delays within each condition, an one-way analysis of variance (ANOVA) was performed. To ensure significant differences and the results were confirmed using the False Discovery Rate (FDR) test \cite{heesen2015adaptive}. To identify within-group differences, the Hoc-Post test was used. Finally, to determine if there were significant differences between predictable and unpredictable conditions for each delay, paired t-tests were conducted.

\section{Findings and Discussion}
The analysis utilized data from 29 individuals, and all the data were examined in both predictable and unpredictable conditions. This comprehensive analysis allowed for a more thorough and accurate exploration of the delays associated with brain connectivity.
\subsection{Comparison of stimulation with clockwise and counterclockwise shapes}
In this section, paired t-tests were conducted to compare the stimulation with clockwise and counterclockwise patterns. For each delay, a separate test was performed for the stimulation with clockwise and counterclockwise patterns across all frequency bands. The results of these tests did not reveal any significant differences. Based on the collected data and examination of the responses, it was observed that approximately 83\% of the given responses were accurate in identifying the clockwise and counterclockwise stimulation.
\subsection{Comparison of delays in two predictability and unpredictability conditions}
In this section, the focus was on examining significant differences between different delays in each separate predictable and unpredictable condition. The statistical test used for this purpose was ANOVA, which revealed significant differences in delays between the two conditions. table\ref{table.1} and table\ref{table.2} display the number of connections with significant differences. The largest differences were observed in the beta, theta, and gamma frequency bands, particularly in the beta and theta bands where the differences were quite noticeable, ranging from 83 milliseconds to 800 milliseconds. These differences were minimal in the delta band and very slight in the alpha band. These differences were observed in both the predictable and unpredictable conditions, although there were more differences in the predictable condition compared to the unpredictable condition. Further details regarding the differences between each delay in the 5 frequency bands are provided below.
\begin{table}[tb]
        \caption{-The number of significant correlations
        $P\text{-Value} \geq 0.05$ 05  among the delays in the predictable condition}
	\centering
	\includegraphics[width= 1\linewidth]{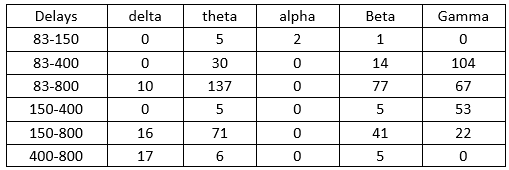}
	\label{table.1}
\end{table}

\begin{table}[tb]
        \caption{-The number of significant correlations
        $P\text{-Value} \geq 0.05$  among the delays in the unpredictable condition}
	\centering
	\includegraphics[width= 1\linewidth]{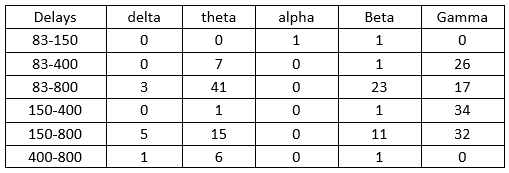}
	\label{table.2}
\end{table}

According to table\ref{table.1} and table\ref{table.2}, in the delta band of the predictable condition, the differences between delays of 83 ms with 800 ms, 150 ms with 800 ms, and 400 ms with 800 ms have increased. In this condition, the differences between the delays of 400 ms with 800 ms were particularly significant, with the majority of connections observed in the frontal region. In the unpredictable condition, the largest difference was observed between the delays of 150 ms and 800 ms, with more connections formed in the posterior region. The significant difference in this condition is relatively smaller compared to the predictable condition.
In the theta band, the highest significant differences between delays of 83 ms with 800 ms, 150 ms with 800 ms, and 83 ms with 400 ms were observed in both the predictable and unpredictable conditions. In the predictable condition, the majority of connections were initially observed between the delays of 83 ms with 800 ms across the entire scalp. Subsequently, the differences between the delays of 150 ms with 800 ms were more prominent, with most connections observed in the frontal region and then in the posterior region. In the unpredictable condition, the largest differences were observed between the delays of 83 ms with800 ms, with most connections seen in the frontal region and then in the posterior region.
In the alpha band, there was a very small but significant difference between the delays in the predictable condition. Two significant differences were observed between 83 and 150 milliseconds, indicating a correlation between the frontal and parietal regions. In the unpredictable condition as well, a significant difference was observed between 83 and 150 milliseconds, indicating a correlation between the parietal and frontal regions. In the beta band, the greatest significant differences in delays were observed between 83 and 800 milliseconds, and 150 and 800 milliseconds, in both the predictable and unpredictable conditions. In the predictable condition, initially the greatest delay was observed between 83 and 800 milliseconds, which was consistently observed throughout the experiment. Subsequently, the differences between 150 and 800 milliseconds were greater, indicating stronger correlations in the parietal region initially, and then in the frontal region. In the gamma band, in the predictable condition, the greatest differences were observed initially between 83 and 400 milliseconds, followed by differences between 83 and 800 milliseconds, with the strongest correlations initially observed in the parietal region and then in the posterior region. In the unpredictable condition, significant differences were observed between 150 and 400 milliseconds, with stronger correlations formed in the posterior region. After that, greater differences were observed between 150 and 800 milliseconds, with stronger correlations in the parietal region.
\subsection{Comparison of Predictable and Unpredictable Scenarios for Each Delay}
In this section, we examined whether there are significant differences between the predictable and unpredictable scenarios for each delay. The analysis was conducted across a frequency range of 1 to 40 Hz, and the network of connections was presented in the following figures. Subsequently, separate calculations were performed for each frequency band. The network of connections in the alpha band was compared to the results presented in reference \cite{samaha2015top}. The results for other frequency bands were also provided along with their analysis. 

To display the results, threshold values were determined based on the average ± standard deviation. If the difference exceeded this threshold, it was considered significant and displayed; otherwise, it was not shown. The thickness of the lines in the figures represents the strength of the connection. 

In the subsequent analysis, the differences between the predictable and unpredictable scenarios were examined for each delay in all frequency bands, and the results are presented below.
\begin{figure}[tb]
	\centering
	\includegraphics[width= 1\linewidth]{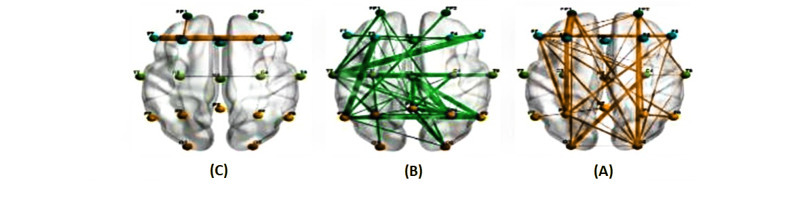}
	\caption{The average correlations in two conditions, predictable (A) and unpredictable (B), along with their significant differences (C), at a frequency range of 1 to 40 Hz for a delay of 83 milliseconds estimated using the PLI method.    
 ( The green line represents a higher average correlation in the unpredictable condition compared to the predictable condition, while the orange line indicates the opposite trend.)}
	\label{fig.3}
\end{figure}

\begin{figure}[tb]
	\centering
	\includegraphics[width= 1\linewidth]{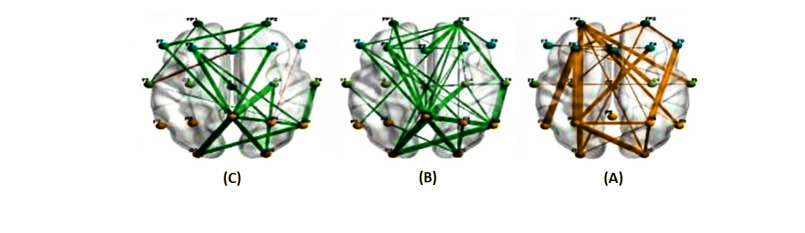}
	\caption{The average correlations in two conditions, predictable (A) and unpredictable (B), along with their significant differences (C), at a frequency range of 1 to 40 Hz for a delay of 150 milliseconds estimated using the PLI method.         
 ( The green line represents a higher average correlation in the unpredictable condition compared to the predictable condition, while the orange line indicates the opposite trend.)}
	\label{fig.4}
\end{figure}

\begin{figure}[tb]
	\centering
	\includegraphics[width= 1\linewidth]{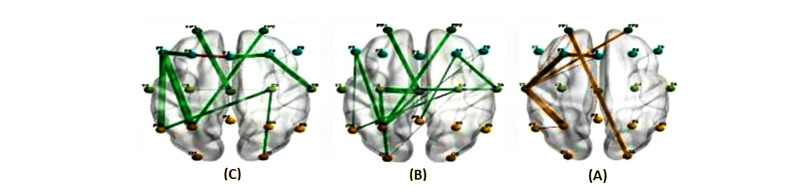}
	\caption{The average correlations in two conditions, predictable (A) and unpredictable (B), along with their significant differences (C), at a frequency range of 1 to 40 Hz for a delay of 400 milliseconds estimated using the PLI method . 
 (The green line represents a higher average correlation in the unpredictable condition compared to the predictable condition, while the orange line indicates the opposite trend.)}
	\label{fig.5}
\end{figure}

\begin{figure}[tb]
	\centering
	\includegraphics[width= 1\linewidth]{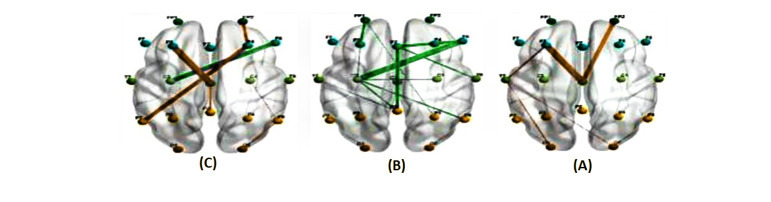}
	\caption{The average correlations in two conditions, predictable (A) and unpredictable (B), along with their significant differences (C), at a frequency range of 1 to 40 Hz for a delay of 800 milliseconds estimated using the PLI method.   
 (The green line represents a higher average correlation in the unpredictable condition compared to the predictable condition, while the orange line indicates the opposite trend.) }
	\label{fig.6}
\end{figure}

\begin{figure}[tb]
	\centering
	\includegraphics[width= 1\linewidth]{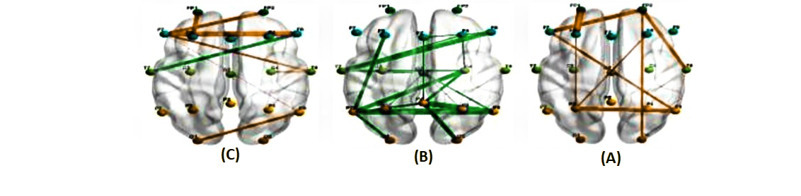}
	\caption{The average correlations in two conditions, predictable (A) and unpredictable (B), along with their significant differences (C), are shown for the alpha band at a delay of 83 milliseconds estimated using the PLI method.
 ( The green line represents a higher average correlation in the unpredictable condition compared to the predictable condition, while the orange line indicates the opposite trend.)}
	\label{fig.7}
\end{figure}

\begin{figure}[tb]
	\centering
	\includegraphics[width= 1\linewidth]{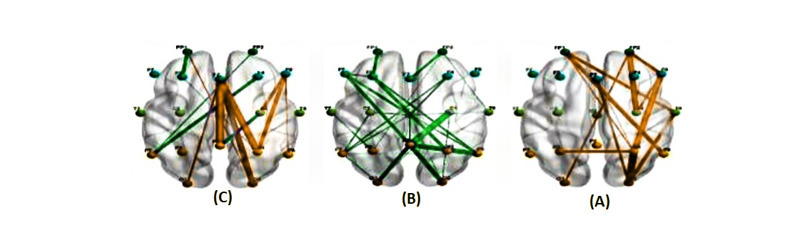}
	\caption{The average correlations in two conditions, predictable (A) and unpredictable (B), along with their significant differences (C), for the alpha band at a delay of 150 milliseconds estimated using the PLI method.        (The green line represents a higher average correlation in the unpredictable condition compared to the predictable condition, while the orange line indicates the opposite trend.)}
	\label{fig.8}
\end{figure}

\begin{figure}[tb]
	\centering
	\includegraphics[width= 1\linewidth]{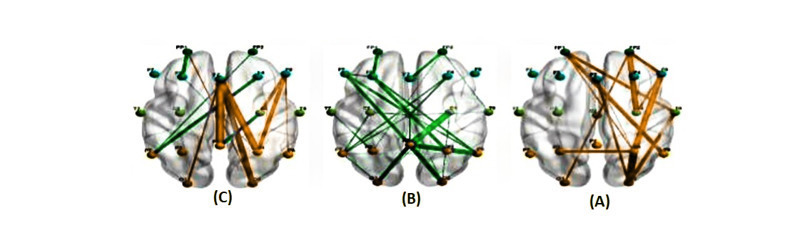}
	\caption{The average correlations in two conditions, predictable (A) and unpredictable (B), along with their significant differences (C), for the alpha band at a delay of 400 milliseconds estimated using the PLI method.        (The green line represents a higher average correlation in the unpredictable condition compared to the predictable condition, while the orange line indicates the opposite trend.)}
	\label{fig.9}
\end{figure}

\begin{figure}[tb]
	\centering
	\includegraphics[width= 1\linewidth]{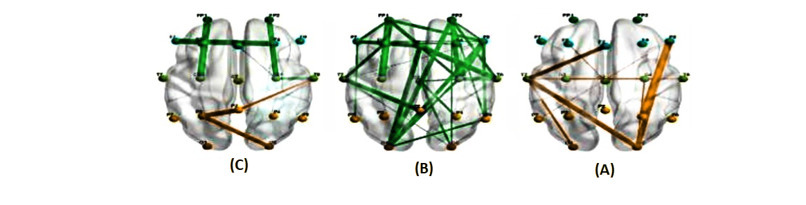}
	\caption{illustrates the average correlations in two conditions, predictable (A) and unpredictable (B), along with their significant differences (C), for the alpha band at a delay of 800 milliseconds estimated using the PLI method.                    
 (The green line indicates a higher average correlation in the unpredictable condition compared to the predictable condition, while the orange line indicates the opposite trend.)}
	\label{fig.10}
\end{figure}

As observed in Figs. \ref{fig.3}--\ref{fig.6}, in the frequency range of 1 to 40 Hz, the connections decrease with increasing delays in the predictable scenario. Additionally, the connections are stronger in the delay of 150 milliseconds compared to 83 milliseconds. In all delays, there is a connection between the two hemispheres. Furthermore, in the delay of 400 milliseconds, a strong connection is observed from the prefrontal region to the posterior region. This connection also exists in the delays of 83 and 150 milliseconds but is weaker compared to the connection in the delay of 400 milliseconds. In the unpredictable scenario there are connections between the two hemispheres. Moreover the connections also decrease with increasing delays. In the delay of 150 milliseconds, strong connections are observed in the parietal and posterior regions compared to other connections at this delay.
To assess the significant difference between the predictable and unpredictable scenarios, a paired t-test with a significance level of $\alpha$=0.05 has been conducted. The results indicate a significant difference in all delays, and particularly, larger significant differences are observed in the delays of 150 and 400 milliseconds compared to the other delays.
in Figs. \ref{fig.7}--\ref{fig.10}, the network of communication in the alpha band is observed, indicating information flow between the posterior and anterior regions. Furthermore, in both scenarios, there is communication between the two hemispheres for all delay groups. It is expected that in the predictable scenario, different regions of the brain are engaged in training at different times. Additionally, the study's findings reveal learning in the alpha band in the parietal regions, even before the stimulation, when the brain is in a training state, aligning with the results of reference \cite{beste2013learning}. Moreover, a study on neural imaging of time perception, described in reference \cite{wittmann2010accumulation}, demonstrated that cortical areas such as the parietal cortex are involved in regulating temporal durations.In the unpredictable scenario, in the 400 ms dela while in the 800 ms delay, there is increased connectivity between the posterior regions and the anterior regions. The findings of this study exhibit differences in brain maps for different delays. To investigate whether there is a significant difference between the two scenarios, a paired t-test with $\alpha$=0.05 has been conducted, and significant correlations have been observed in all delays. In the 400 ms delay, the average connectivity in the unpredictable scenario is higher than in the predictable scenario. In the 800 ms delay, the average differences in connectivity between the unpredictable and predictable scenarios are greater in the anterior regions compared to the posterior regions. Similar significant differences have been observed in the other frequency bands, which are reported in detail below.
In the delta band, for the 150 ms, 400 ms, and 800 ms delays in the predictable scenario, there was connectivity between the central and frontal regions. Additionally, in the 83 ms delay, there was strong connectivity from the central region to the anterior region. Moreover, in the 800 ms delay, there was a very strong connection from the central region to the frontal region. In the unpredictable scenario, for all delays, the number of connections was higher compared to the predictable scenario. In this scenario, although there were strong connections in the 800 ms delay, the number of connections decreased compared to other delays. In the 83 ms and 150 ms delays, there was also a very strong connection between the parietal and frontal regions.
In the theta band, in both scenarios, there were connections observed in the frontal, prefrontal, and parietal regions for all delays. However, in the 800 ms delay, the number of connections decreased compared to other delays. Additionally, in the study mentioned in reference \cite{fontes2020time}, it has been shown that exposure to timing estimation tasks affects the severity of inattentive symptoms, ADHD, and theta band activity in the lateral posterior frontal cortex.
In the beta band, in the 800 ms delay, there was a decrease in connections compared to other delays in both scenarios. Additionally, the prefrontal region had connections to other areas in all delays, and in the unpredictable scenario, there were more connections compared to the predictable scenario.
In the gamma band, in the unpredictable scenario, there were connections throughout the brain for all delays, and in the predictable scenario, there were connections for the delays of 83 and 150 milliseconds. Additionally, there were connections between the posterior and anterior regions in both scenarios. Furthermore, in the article referenced as \cite{mioni2020understanding}, it is shown that intervals of less than a second and processing time are associated with increased cognitive functions in areas including the prefrontal cortex.
The results reported in paper \cite{samaha2015top} indicate that predictability has an effect on behavior at a delay of 800 milliseconds. The authors concluded that there is a predictive timing effect on alpha power during the longest delay period (800 milliseconds). In their experiments, it was found that after the presentation of a cue, alpha power decreased in the predictable condition (between 172-373 milliseconds after the cue) compared to the unpredictable condition, while before the stimulus presentation, alpha power increased in the predictable condition (305 milliseconds before the stimulus) compared to the unpredictable condition. Their study only observed a significant difference between the predictable and unpredictable conditions at the longest delay (800 milliseconds), whereas, in the current study, a significant difference between the two conditions was observed at all delays.
\section{conclusion}
The results indicate that there was no significant difference observed between the clockwise and counterclockwise stimuli. Comparing the delays in both the predictable and unpredictable conditions, the findings suggest that the greatest differences in delays were observed in the gamma, beta, and theta bands, with the largest differences between the 83 and 800 milliseconds delays in the beta and theta bands. In the study by Rajkumar et al. \cite{rajkumar2020bio}, it was shown that participants who perceived time as shorter than the physical time exhibited higher beta power and higher coherence in central regions. Additionally, in the study by \cite{rivera2020correlation}, a correlation matrix of connectivity was demonstrated between the frontal-central and posterior regions across all frequencies. In the current study, it was observed that, in comparison of delays, both in the predictable and unpredictable conditions, there were connectivity patterns between the frontal-central regions in all frequency bands except for alpha.
In this section, a significant difference was observed between the delays, with a greater difference observed in the predictable condition compared to the unpredictable condition. By examining these significant differences, it is possible to extract the regions that were more involved in the process.
\begin{table}[tb]
        \caption{The greatest differences have occurred between the delays and the involved regions in both the predictable and unpredictable states}
	\centering
	\includegraphics[width= 1\linewidth]{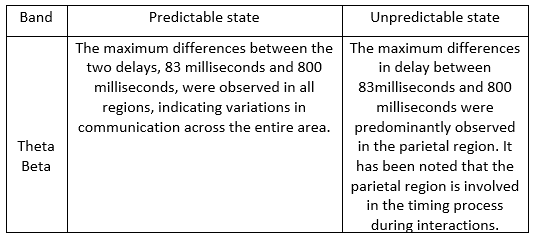}
	\label{table.3}
\end{table}

\begin{table}[tb]
        \caption{Comparison between the predictable and unpredictable state}
	\centering
	\includegraphics[width= 1\linewidth]{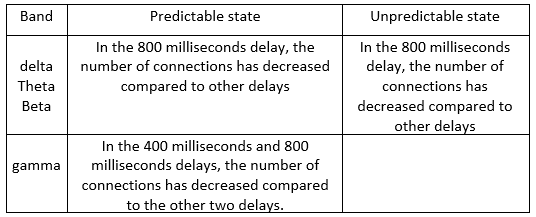}
	\label{table.4}
\end{table}

The article referenced in \cite{merchant2013interval} states that the brain does not exhibit significantly different neuronal spike activity in time intervals less than 500 milliseconds. Therefore, it is suggested to design an experiment with delays ranging from 2 to 3 intervals above 500 milliseconds in order to analyze temporal events with greater clarity and accuracy. Additionally, to reduce participant fatigue without compromising the integrity of the experiment, it is recommended to use fewer trials per delay group.

\bibliographystyle{IEEEtran}
\bibliography{IEEEabrv,mybibfile}

\end{document}